# "Read My Lips": Using Automatic Text Analysis to Classify Politicians by Party and Ideology[1]

Eitan Sapiro-Gheiler[2]

June 15, 2018

Department of Economics

Princeton University

[1] Acknowledgements: I would like to thank my advisor, Professor Adrien Matray, for all his insights and support throughout this project, and Professor Silvia Weyerbrock, for all her help with the junior independent work program.

[2] E-mail address: eitans@princeton.edu

**Abstract**

The increasing digitization of political speech has opened the door to studying a new dimension of political behavior using text analysis. This work investigates the value of word-level statistical data from the US Congressional Record—which contains the full text of all speeches made in the US Congress—for studying the ideological positions and behavior of senators. Applying machine learning techniques, we use this data to automatically classify senators according to party, obtaining accuracy in the 70-95% range depending on the specific method used. We also show that using text to predict DW-NOMINATE scores, a common proxy for ideology, does not improve upon these already-successful results. This classification deteriorates when applied to text from sessions of Congress that are four or more years removed from the training set, pointing to a need on the part of voters to dynamically update the heuristics they use to evaluate party based on political speech. Text-based predictions are less accurate than those based on voting behavior, supporting the theory that roll-call votes represent greater commitment on the part of politicians and are thus a more accurate reflection of their ideological preferences. However, the overall success of the machine learning approaches studied here demonstrates that political speeches are highly predictive of partisan affiliation. In addition to these findings, this work also introduces the computational tools and methods relevant to the use of political speech data.

# 1 Introduction

It is commonly believed that politicians lie, and do it often. However, existing political science literature on promise-keeping is more mixed than that adage would suggest. Both on specific issues and generally, political parties are surprisingly trustworthy, e.g., Pétry and Collette (2009). Early work in this area manually examined party platforms and similar documents to derive a list of promises and cross-referenced them with policy implementations, as in Bradley (1969) or Budge and Hofferbert (1990). Such analysis even entered the journalistic mainstream–PolitiFact kept an "Obameter" hand-tracking President Obama's promise-keeping (http://www.politifact.com/truth-o-meter/promises/obameter/). Automated text analysis techniques provide a new set of tools with which to address these questions. Recently, Lauderdale and Herzog (2016) and Gentzkow et al. (2016) have used these modern methods to quantify political polarization by extracting features from speeches given in the US Congress. The work presented here builds on these early contributions by examining the value of political speech for predicting partisan affiliation. This approach allows us to directly test whether politicians' speech, and not just their votes, is an accurate indication of their policy positions.

A key theoretical difference between speeches and votes is that they represent two different levels of commitment. While speeches are public, they are nonbinding and are often treated as poorly predictive. On the other hand, votes, more specifically US Congressional roll-call votes, represent recorded positions taken on legislation that have an influence on policy. These votes can generally be treated as representing politicians' revealed preferences because of their higher commitment cost. However, given that many votes pass by large margins, the possibility of performative votes–those taken to showcase a position, not influence the outcome–mean that roll-call votes are not necessarily a perfect measure of true preferences. Moreover, there may be selection bias in terms of which topics are brought to a roll-call vote, as discussed by Carrubba et al. (2006). Automated text analysis, as discussed in this work, provides a new paradigm to address these challenges.

By using methods from machine learning to predict political party from political texts, this work achieves two ends. First, since party membership is recorded and objective, it validates the accuracy of various classification methods. Second, classification by party replicates, in a simplified sense, voters' experience, since voters attempt to determine a candidate's ideology from available information. While party-blind elections are uncommon, they do exist, especially for judicial nominees, e.g., Bonneau and Cann (2013) and Burnett and Tiede (2014). Furthermore, primaries can be seen as "party-blind" elections in which voters face the task of choosing the more liberal or conservative candidate without party labels. Fitting the same classification models to ideology rather than party provides a more accurate simulation, though without the direct verifiability of partisan classification.

This work shows that the text of political speeches is indeed highly predictive of party affiliation. Using four types of machine learning models, we achieve classification accuracy in the 70%-95% range depending on the model and the data used to train it. These successful results lead to three main conclusions. First, they show that political speeches are valuable for determining politicians' broad ideological positions. Second, they provide large-scale quantitative support for earlier works which manually analyzed the concordance of political speech and political action. Finally, the differing accuracies of the models used here clarifies which directions of refinement are likely to be most successful in providing even stronger results regarding classification or ideology identification from political speech.

Section 2 summarizes closely related literature. Section 3 describes the data used for this work, and Section 4 presents the text analysis approaches. Section 5 analyzes numerical results and Section 6 provides conclusions and directions for future research. A data excerpt and description of data preparation are included in Appendix A; details of the machine learning methods used are in Appendix B; and tables and figures appear in Appendix C.

## 2 Related Literature

There is ample literature on the trustworthiness of political parties. On the theoretical side, Aragonès et al. (2007) develop a model in which purely ideological candidates adhere to campaign promises under threat of punishment by voters. This argument holds up empirically; Pétry and Collette (2009) find in their meta-analysis of studies on political trustworthiness that parties keep 67% of their promises, albeit with wide variation. More germane to this paper are specific analyses using party-platform data to analyze trustworthiness in the US. Bradley (1969), Royed and Borelli (1997), and Budge and Hofferbert (1990) all manually check promise-keeping in political documents and argue that most salient promises are kept. Elling (1979) argues that differences in the language party platforms use are also predictive of trustworthiness. Despite some ambivalence as to the degree of the effect, there is broad agreement that party platforms predict political action.

Because they are recorded and publicly available, roll-call votes have become the primary basis for voting-based measures of political ideology. However, there is some literature on the inaccuracy of these votes as predictive measures. Hug (2010) argues that there is selection bias in which votes are recorded as roll-call votes in the Swiss Parliament and Carrubba et al. (2006) made a similar argument using European Parliament data, though they find a weaker selection bias. These works note, however, that in US Congress every vote is a recorded roll-call vote, so selection bias is vastly less prominent if it is present at all.

In its raw form, roll-call voting behavior does not provide a clear picture of individual politicians' revealed preferences. Poole and Rosenthal (1985) provide the seminal work in the field of aggregating these votes to generate numerical estimates of political ideology. Their work generates what are known as Poole-Rosenthal DW-NOMINATE scores, henceforth DWN scores, the standard in the economics and political science literature. DWN scores are based on the idea that politicians' utility functions provide them with an "ideal point" in Euclidean space that describes their preferences and that politicians with more similar voting records have closer ideal points.

The treatment of political speeches as data composed of words rather than units of discourse is a relatively new approach. Laver et al. (2003) developed the pioneering Wordscores model to compare texts with unknown positions to those with a priori known positions. Slapin and Proksch (2008) develop Wordfish, a time-series version of the Wordscores model. These works use a variety of text sources; however, due to its size and richness, the US Congressional Record has become common both as a testing ground for new models and as a topic of research in and of itself. For instance, Quinn et al. (2010) determine clusters of speeches that are about related issues, find words that characterize each cluster, and predict speeches' location in these clusters. Jensen et al. (2012) follow a similar approach, extracting particularly partisan "trigrams," or sets of three words, then using the frequency of these phrases to quantify partisanship. Lauderdale and Herzog (2016)'s analysis is the closest to the goals of this paper. Using data from the Irish Daíl and US Congressional Record, they first divide the text into "debates," or sets of speeches on particular bills. They then use the Wordfish algorithm to determine the ideological position of speakers in each individual debate and determine ideological positions based on a common scaling of the Wordfish scores from each debate. Gentzkow et al. (2016), like Jensen et al., look at trends in political polarization, but, to reduce finite-sample bias, use a Bayes' Rule-based model for two-word "bigrams" to assign the probability that speakers belong to a given party.

Most recent work in text-as-data analysis of political speeches has a two-pronged goal. First, progress in mathematics and computer science is used to develop models that more accurately classify or describe text in ways that align with observed phenomena, such as political polarization. Second, these new models are used to draw conclusions about these real-world phenomena by performing analyses that exploit the fine-grained, quantitative measures of polarization or ideology recovered from the model. This work focuses on that second goal by using machine learning models for inspiration and examining the predictive value of political speech directly rather than as a consequence of model selection.

## 3   Data

To construct a measure of politicians' positions based on statements they have made, we employ automatic text analysis methods to parse a selection of speeches. Both text analysis and classification require a large text base–referred to as a "corpus"–to be effective. We use data compiled by Laurendale and Herzog (2016) from the Congressional Record for the 104th through 113th Senate (1995-2014); see Appendix A for examples. The Congressional Record included all speeches made in the House of Representatives or Senate during a congressional session, and as such, is a reasonable corpus of political speeches. The authors note that the Congressional Record may be amended, but amendments are minor and do not substantially alter the content of the text. 1,009 unique senator-sessions—pairs of a senator's name and a session of Congress, used to treat the same senator serving in multiple sessions as multiple individuals—are available. For some senators, other data is not available, so they are dropped from the relevant sections of the analysis where that is the case. In all cases, the sample contains at least 1,000 senator-sessions.

As discussed in Section 2, in the economics and political science literature, DWN scores, drawn from voteview.com, are the standard aggregations of Congressional roll-call votes into numeric estimates of political ideology. The two dimensions of these scores represent economic/redistributive policy and social/racial policy. For the time period we are studying, the first dimension is vastly more predictive and is thus the focus of our analysis. Since the goal is to assess voters' preferences at each election opportunity, each term a legislator served is treated as a data point. However, DWN scores are fixed for individual legislators across their careers to allow for comparison between sessions of Congress, so the same DWN scores are used for the same senator across different senator-sessions.

# 4 Methodology

## 4.1 Text Processing

Before attempting to classify senators, the corpus needs to be cleaned so that it is usable. This cleaning is described in Appendix A. Briefly, all but English words are removed and stemming is applied to convert words with different conjugations to be the same word stem, e.g., "vote," "votes," "voted," would all become "vote." The standard algorithm for doing so comes from Porter (1980), used for example by both Gentzkow et al. (2016) and Lauderdale and Herzog (2016). The first discretionary step is choosing "relevant words." To do so, we sort words by number of appearances, deleting those with too many or too few. We use only word stems appearing over 1,000 times, or an average of one use per senator-session in our sample. These cutoffs are intended to eliminate words that may be highly predictive but only in virtue of their extremely narrow usage. Thus, the decision to exclude them is not only a technical choice but also a reflection of our intuition that voters are unable to scrutinize every single word a politician has ever said, only those that are reasonably frequent. Histograms in Appendix A show that this cutoff does not meaningfully alter the data.

## 4.2 Party-Based Classification

Once the corpus has been pre-processed, speeches are assigned to senator-sessions. While a senator serves in three sessions of Congress before being reelected, congressional priorities change from session to session due to the reelection of House members, so this fine-grained analysis is preferable. We take a data-driven approach to classification and implement machine learning methods rather than developing intuitive but potentially biased approaches. Four classification methods are used: decision tree, naïve Bayes, support vector machine (SVM), and lasso-penalty regression. Details of all four methods are in Appendix B.

All classification methods used in this paper are based on 10-fold cross-validation, in which the algorithm partitions the data, uses the data minus one partition element to train the classifier, tests on the excluded partition element, and then repeats this process for all partition elements. We train each method using all ten sessions of Congress and test its performance on both the full sample and each individual session. We also train classifiers using only the data from each individual session of Congress, resulting in 10 more classifiers of each type. Each of these is then tested not only on its own session, but also on true "out-of-sample" data by applying it to each of the other sessions of Congress and the combined 10-session sample. This results in four 11-by-11 matrices, which include the accuracy rates of the overall and Congress-specific classifiers for each classification method across the full sample and each of the 10 individual sessions of Congress; see Table 2 in Appendix C.

## 4.3 DWN-Based Classification

We can refine our classification strategy by relying on a finer target variable, namely DWN first-dimension scores, henceforth DWN1 scores. These scores are normally distributed within each party but the two distributions overlap minimally. Classifying senators with scores left of the midpoint as Democrats and those with scores right of the midpoint as conservatives has an accuracy rate of 99.11% in our sample. The second-dimension DWN2 scores are also normally distributed within each party, but the two distributions are almost entirely overlapping, so a similar method has only 45.19% accuracy in our sample. If we take the DWN1 score as a "true" measure of ideology, this is an even closer approximation of the task voters face, as they attempt to assign an ideological position, not just a party affiliation, to politicians whose stances they do not yet know.

Because this task is more complex than binary classification, we adjust our classification approaches accordingly. Tree classifiers and SVMs have natural analogues that can predict continuous variables, so we implement those methods. Standard naïve Bayes classifiers rely on sorting data into a discrete number of classes, so we consider two possible approaches to adapt it to this task. One would be to define classes given by bins, e.g., of width 0.1, ranging from a score of -1 to a score of 1. However, this approach has challenges with bins where no data points exist, a problem that is exacerbated when the bin size decreases to allow for finer classification. Alternatively, the DWN1 scores could be used as prior probabilities of belonging to one party or another. Because each party's scores are approximately given by a normal distribution, we can compute the probability that a given score is drawn from one of the two distributions and use this as a prior. However, the standard deviation of these fitted distributions is not large enough to yield priors significantly different from 0 or 1, even if we fit fat-tailed distributions instead of normal distributions. A 0/1 prior renders classification moot. As such, we choose not to include a naïve Bayes analogue for this portion of the analysis, as it does not significantly affect the conclusions of this work. Finally, adapting the lasso model is straightforward. Following the observed distributions, we fit a lasso model to each party separately and assume a normally distributed outcome variable.

There remains the question of validating the accuracy of these DWN1 predictions; since the DWN1 scores are continuous, requiring that the prediction match the DWN1 score is not an option, and requiring a match within a given range leaves open the question of what level of error is natural. As such, we opt to validate in a manner that allows comparison to the results of Section 4.2. We label senators with predicted DWN1 scores of less than 0 to be Democrats and senators with predicted DWN1 scores of greater than 0 to be Republicans, then compute the accuracy rate of this assignment. Table 3 in Appendix C shows the results of this validation in a manner similar to Table 2 in the same appendix.

## 5  Discussion of Numerical Results

While all methods show the power of text for predicting party affiliation, the full-sample lasso classifier outperforms the other full-sample classification methods. The top-left entry of each matrix in Table 2 gives the accuracy rate across the full sample. The four classifiers have accuracy rates of 74.53%, 72.75%, 89.99%, and 98.32% for tree, naïve Bayes, SVM, and lasso respectively. The mean single-session classifiers have accuracy rates of 65.07%, 72.89%, 76.08%, and 67.85%, showing that the lasso does not perform as well across the full sample with limited data. However, it tends to outperform the others in sessions of Congress more distant from the one used to train the classifier. This result should be interpreted with caution. Cross-validation performed on the tree, SVM, and naïve Bayes classifiers ensures that training data is never used for testing by partitioning data into $n$ bins and returning $n$ sub-classifiers, each valid for $1/n$ of the data; in this work, we use $n = 10$. The overall classifier's performance is computed as the sum of correct predictions made by each sub-classifier within its relevant bin for within-sample tests and the average of the sub-classifiers' performances for out-of-sample tests. In contrast, the lasso classifier uses cross-validation to set the $\lambda$ parameter in the lasso formula (Equation (2) in Appendix B). The algorithm then outputs one set of coefficients to be used for all data points, including those that were used as training data in some of the cross-validated attempts. Thus, a reasonable approximation for the lasso performance within-sample would be to consider by how much the lasso outperforms a rate of 0.9—equivalent to successfully fitting all the training data—and scaling that up by 10. This is a lower bound, as the lasso algorithm explicitly avoids over-fitting and thus the resulting classifier may "intentionally" misclassify some training data. Under this specification, the lasso still tends to outperform the other classifiers, albeit with the caveat that this rough approach is not apt for precise comparisons.

Across methods, there is a rough decrease in performance as the test data is drawn from sessions further away from the training data, indicating evolving trends in partisan speech. The naïve Bayes and SVM classifiers have least variance in their accuracy rate. We can measure this by computing the standard deviation of the accuracy rate among the single-session classifiers of each type, then taking the average of those standard deviations. The tree has a value of 0.1972 compared to 0.0848 for the SVM model and 0.0954 for naïve Bayes model. However, the naïve Bayes classifier does somewhat worse on nearby sessions than the others, perhaps indicating that the assumption of independence of each word's distribution (discussed in Equation (1) in Appendix B) is violated. Both the naïve Bayes and SVM classifiers tend to perform slightly worse in-sample than in nearby sessions. This phenomenon has a straightforward explanation. In-sample testing is done so that no partition classifies its own training data, ensuring out-of-sample behavior. In neighboring sessions, even senators used for training are considered "out-of-sample," so a partition trained on a senator in, e.g., the 105th session counts accurate classification of that senator in the 106th session. This "quasi-in-sample" effect works in the opposite direction of the divergence of vocabulary over time. Thus, for the naïve Bayes and SVM classifiers, the quasi-in-sample effect dominates because the drop-off in performance over time is not as steep. This mathematical effect provides support for two intuitive claims: first, that senators' vocabulary is more constant during the terms they serve than the vocabulary of Congress at large, and second, that over time, Congressional language changes both among and between parties.

Also of interest are occasions on which classifiers yield a result less than 0.5, indicating that predicting the opposite of what a classifier suggests would be more accurate than chance, i.e., the classifier is "backwards." While this could be a technical result related to the particularities of the classifier, it occurs across three different classification methods and thus that is unlikely to always be the case. This issue is most common for the classification tree, which relies on binary assessments. It occurs 42 times compared to 6 for the SVM, 3 for the naïve Bayes model, and 15 for the lasso model. We therefore propose a reason for this empirical observation: certain words may "change party" over time. For instance, it seems intuitive that words like "budget" or "pass" (as in, pass a bill) would be associated with the majority party, while "veto" might be associated with the minority party, especially if they hold the presidency. The coefficients of the lasso-penalty regression represent the words whose counts are most useful in classification, and thus can be interpreted as representing one measure of word partisanship. Extracting words with the greatest magnitude of change in coefficients between each pair of sessions of Congress would allow detailed analysis of how partisan vocabulary changes between sessions. This analysis is left for future work.

Classification using DWN1 scores as validated by party classification does not have a significant effect on the performance of the tree classifier. The whole-Congress tree performs 2.20% worse on average when trained with DWN1 data and the average individual-session classifier performs 0.85% worse on average. However, no classifier's average performance changes by more than 5%. No session of Congress becomes more than 2% more difficult to classify on average, though no session becomes easier to classify on average. Other than this mild overall decrease in performance, there is no clear pattern regarding the tree-session pairs that result in better or worse accuracy rates.

Comparing the party-trained SVM to the DWN1-trained SVM, the whole-Congress SVM performs markedly worse when trained with DWN1 data, with an accuracy rate about 20% lower in each Congress and overall. However, the individual-Congress SVMs show no clear change in average performance. The average change in average performance is a 0.33% decline in accuracy, with individual changes ranging from a 6.33% average improvement in the 111th Congress SVM to a 5.50% decline in performance in the 112th Congress SVM. No session of Congress becomes more than 3% easier or more difficult to classify across all individual-Congress SVMs. Also of note is the absence of a clear pattern as to which SVM-session pairs result in better or worse accuracy rates. These results point to the difficulty faced by the whole-Congress SVM being a result of the aggregate variation of DWN1 scores rather than any given Congress being difficult to classify.

These results suggest that introducing the intermediate step of determining senator ideology before assigning party classifications does not provide any clear benefit for classification, and may in fact serve as a hindrance. However, the lasso classifier shows precisely the opposite result, rising from its already-impressive average performance of around 70% accuracy to near-perfect classification when this intermediate step is introduced. Indeed, the lasso's performance only falls below 90% in three lasso-session pairs, with the lasso generated from the 106th Congress achieving accuracy rates of 72.12%, 87.88%, and 86.00% when classifying the 108th, 109th, and 110th Congresses respectively. This markedly different result clarifies that the DWN1 scores are not uninformative, but that their informational value depends on the model through which they are interpreted. Because the lasso model achieves precise estimates of the DWN1 scores, it is then able to leverage those estimates into extremely accurate classification. The lasso model's errors in estimating DWN1 scores are on the order of 0.01 or less; this means only those scores that are already close to 0, or those senators whose scores cross the ideological midpoint, will lead to misclassification. In contrast, the larger error rates in score prediction exhibited by the tree and SVM classifiers mean that the DWN1 scores may not be more valuable than directly classification through party labels.

## 6 Conclusion

This work applies automatic text classification models to the question of party identity. In its simplest form, this is the task that voters face at the ballot box. Understanding whether and how this behavior can be replicated computationally provides insight into voters' ability to sort candidates by ideology. We show that political speech data is a powerful tool for predicting partisan affiliation, with the most accurate models consistently achieving accuracy rates around or above 90% and all models consistently achieving accuracy rates above 70%. Predicting DW-NOMINATE scores instead of party affiliation and then extrapolating party affiliation from those predictions has mixed results. While the lasso model improves significantly in accuracy, the tree and SVM models perform slightly worse. These results imply that some models may be more effective tools for this data than others, and that the choice of model ought to reflect the specific task it is intended for. Overall, all the results here show that accurate identification of party from political text is both possible and probable. The decreasing effectiveness of classification as the data to be classified moves further away in time from the training data is an important result showing that temporal variation is present in Congressional speech. This conclusion indicates that voters who do not keep up with political developments will likely lose the ability to perform the basic partisan-sorting task after only one or two sessions of Congress.

There are three primary limitations to this analysis, which point to directions for future work. The first is the classification models used, which do not cover the range of machine learning techniques. Most notably, neural network approaches, e.g., Goodfellow et al. (2017), were not considered, in large part due to the volume of data needed for training. However, refinements of these methods, e.g., transfer learning, have the potential to further improve upon the already-successful results of the classification in this work. Secondly, there is room for improvement with respect to the datasets used. While the corpus drawn from the Congressional Record is extensive and thorough, it may not be a fully accurate representation of the public image of politicians. This provides ground for further research that uses web-scraping and other techniques to perform similar text-based analysis using public speeches and statements, providing an even more accurate picture of outward-facing political behavior. Finally, there are a variety of non-word-count-based methods of text analysis, such as word embedding or natural language processing, e.g., Mikolov et al. (2013), that could incorporate more complex relationships between words and phrases or even include meaning as a component of the analysis. These approaches would further enrich both the predictive power and the verisimilitude of these models as they relate to the task faced by voters. The realm of text analysis is a growing field with great potential for applications in political science, economics, and beyond. This work is a first step towards taking full advantage of these rich new tools and datasets.

# Appendix A: Data Processing

## A1: Data Excerpt

Laurendale and Herzog (2016) present the unmodified Congressional Record and combine speeches by speaker in the first step of processing. An excerpt, from a speech given on April 30, 2002 in the 107th Congress by Fritz Hollings (D–SC) is below:

> 148 51 Senate S3515-S3522 107 2 Tuesday 30 April 2002 ANDEAN TRADE PREFERENCE ACT–MOTION TO PROCEED Mr. HOLLINGS We have to get a value-added tax to pay for this war on terrorism that is costing the country and offset the 17-percent value added tax advantage. For example, in Europe where it is rebated, it is costing us a 17-percent differential in trade right there.
>
> 148 51 Senate S3515-S3522 107 2 Tuesday 30 April 2002 ANDEAN TRADE PREFERENCE ACT–MOTION TO PROCEED Mr. HOLLINGS Enforce our dumping laws, but please do not say you have to get more productive. What is not producing is not the industrial worker in the United States, it is the U.S. Congress. We haven't produced. We have been running around like lemmings: Free trade, free trade, fast track, fast track–having no idea in the Lord's world what we are doing; whereas we are exporting jobs faster than we can create them.

## A2: Data Cleaning

Data cleaning and analysis was performed in Matlab. The key steps in cleaning (with corresponding Matlab functions in parentheses) are: erase punctuation (erasePunctuation), convert to lowercase (lower), create an array of documents (tokenizedDocument), remove common "stop words" (removeWords(stopwords)), remove words with too few or too many characters (removeShortWords, removeLongWords), apply the Porter stemmer from Section 4.1 (normalizeWords), and create a model containing word counts (bagOfWords).

The average Congressional session in this sample, which includes the $104^{th}$-$113^{th}$ Congresses, contains 728,000 speeches, 18,105,690 words, and 136,064 cleaned word stems. Table 1 includes example data on word frequency. Note that the words are stemmed; "senat," for example, includes "Senate," "Senator," "Senators," etc. Also note there may be minor discrepancies (e.g., the sum of word use across Republicans and Democrats may not equal the overall usage of a word) due to some particularities of the data, such as the need to hand-code the party of various Congressional officials or the exclusion of procedural speeches in the cleaned data.

Table 1: Top 10 Words in 104th Senate Speeches

| Overall | | Democrats | | Republicans | |
|---|---|---|---|---|---|
| Words | Count | Words | Count | Words | Count |
| "bill" | 170900 | "senat" | 66034 | "senat" | 69896 |
| "senat" | 166830 | "presid" | 60934 | "presid" | 68968 |
| "presid" | 165150 | "bill" | 42463 | "amend" | 41641 |
| "amend" | 155570 | "amend" | 42316 | "bill" | 40680 |
| "year" | 152690 | "year" | 40100 | "state" | 38813 |
| "state" | 139420 | "state" | 37860 | "year" | 36992 |
| "time" | 128020 | "peopl" | 29642 | "time" | 28840 |
| "peopl" | 122780 | "time" | 28823 | "peopl" | 26559 |
| "speaker" | 100340 | "work" | 25944 | "budget" | 25718 |
| "work" | 96686 | "budget" | 25385 | "think" | 21344 |

Source: Own calculations using data from Laurendale and Herzog (2016).

Figure 1 shows frequency distributions of stemmed words in the full dataset; they are roughly exponential regardless of cutoff. There are 386,599 unique stemmed words; 15,380 are used >100 times; 5,090 are used >1,000 times; and 1,467 are used >10,000 times.

**Figure 1**: Distribution of stemmed words. From left to right: all words, words used >100 times, words used >1,000 times (the cutoff for this work), words used >10,000 times

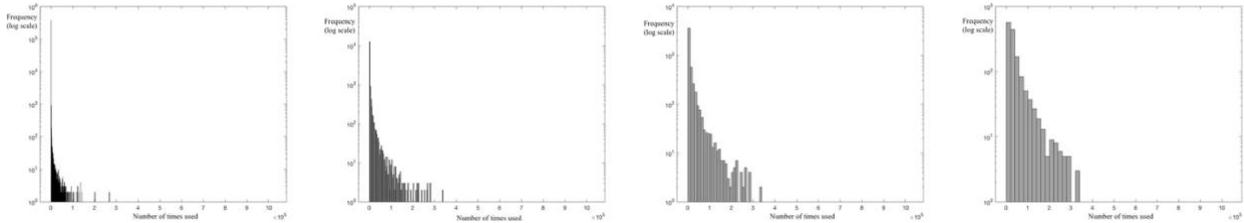

Source: Own calculations using data from Laurendale and Herzog (2016).

## Appendix B: Classification Methods

The first method used is a decision tree classifier. Decision tree classification functions by forming a "tree" of queries, called nodes, that result in binary or numeric responses. Depending on the result of the previous query, a new query is proposed until the data has been sorted. For example, a simple classification tree might ask, "Is 'Iraq' used >100 times?" If yes, it asks, "Is 'tax' used > 65 times?" and if no, it asks "Is 'foreign' used >75 times?" Each data point—here, a word count for a given senator—travels a singular path along the branches. Node choice is determined by an algorithm based around impurity, the degree to which the branches of a node line up with the classes for classification.

The second method is a naïve Bayes classifier. This approach treats each word count as a random variable that, conditional on the speaker being Republican or Democratic, is independent of other word counts. The naïve Bayes classifier estimates the distributions of these random variables and then implements Bayes' Rule:

$$\hat{P}(Y = k \mid X_1, \ldots, X_p) = \frac{\pi(Y=k) \prod_{j=1}^{p} P(X_j | Y=k)}{\sum_{k=1}^{2} [\pi(Y=k) \prod_{j=1}^{p} P(X_j | Y=k)]} \quad (1)$$

where $Y$ is party affiliation, the $X_j$ are the random variables for word $j$'s count, and $k = 1$ or $2$ is party. $\pi$ is the prior probability of belonging to one of the classes, here the proportion of the total population in a given class. Each word used must be used by enough senators to allow distribution-fitting to take place. As such, about 50 words were dropped; however, none of those words appears in the top 1,000 most common words.

The third classifier studied here is a support vector machine, or SVM. An SVM constructs a set of hypersurfaces in high-dimensional space to separate the testing data. Then it classifies testing data by determining where it lies in the space designated by those hypersurfaces. Specifically, the algorithm defines for an observation $x$ and bias term $b$, the hyperplane $x'\beta+b = 0$ by its orthogonal vector $\beta$. The SVM minimizes $\|\beta\|$ and $b$ subject to $y_i f(x_j) \geq 1$ for all data points $(x_j, y_j)$. If a hyperplane separating the data cannot be found, we introduce additional variables $\xi_j$ to represent the magnitude of any misclassification and $C$ to represent the penalty for these mistakes and minimize $(\beta'\beta)/2 + C \sum_j \xi_j^2$ with the constraints $y_i f(x_j) \geq 1 - \xi_j$ and $\xi_j \geq 0$.

The final classification method used is lasso-penalty regression. Because there are thousands of "predictor variables" in the form of word counts for each word used in the corpus, traditional ordinary least squares regression would perform poorly. Lasso-penalty regression adds a penalty as coefficients become different from 0 by solving

$$\min_{\beta_0, \beta} \left( \frac{1}{N} Dev(\beta_0, \beta) + \lambda \sum_{j=1}^{p} \beta_j \right) \quad (2)$$

where $\beta_0$ is the intercept, $\beta$ is the vector of the coefficients $\beta_j$, $Dev$ represents the deviance (here to the log-likelihood of the classifier fitting the true data), and $\lambda$ is the lasso penalty parameter. This penalty produces a classifier capturing the true effect of the predictor variables, not all of which are likely to be relevant. The value of $Dev$ depends on the choice of "link function," which we choose to represent the distribution of our data. For party classification, we assume a binomial distribution and use a logit link function

log[$\mu/(1 - \mu)$] = $Xb$ where $\mu$ is our dependent variable, $X$ is the matrix of word counts, and $b$ is the vector of coefficients we are trying to find. For DWN1 classification, we assume a normal distribution and use the identity link function $\mu = Xb$. A lasso classifier also generates, via the coefficients with largest magnitude, a set of "most partisan words." We use this list to validate that the words dropped from the naïve Bayes classification due to insufficient data are not particularly partisan.

## Appendix C: Classification Results

Table 2: Accuracy Rate of General and Congress-Specific Classification Using Party Data

| Training Session | Rate Correct by Session of Congress, Tree Classifier | | | | | | | | | |
|---|---|---|---|---|---|---|---|---|---|---|
| | All | 104th | 105th | 106th | 107th | 108th | 109th | 110th | 111th | 112th | 113th |
| All | 0.7453 | 0.7451 | 0.7400 | 0.7822 | 0.6837 | 0.7526 | 0.8200 | 0.7327 | 0.7170 | 0.7200 | 0.7573 |
| 104th | 0.6909 | 0.7941 | 0.8050 | 0.8000 | 0.5173 | 0.7732 | 0.7630 | 0.2960 | 0.3887 | 0.4250 | 0.4359 |
| 105th | 0.6561 | 0.7598 | 0.8500 | 0.8713 | 0.4969 | 0.8072 | 0.8560 | 0.2426 | 0.3321 | 0.2690 | 0.2786 |
| 106th | 0.6672 | 0.8098 | 0.8650 | 0.8119 | 0.5367 | 0.7938 | 0.8480 | 0.2693 | 0.2538 | 0.3780 | 0.3136 |
| 107th | 0.6825 | 0.5716 | 0.6140 | 0.6376 | 0.5612 | 0.5959 | 0.6090 | 0.5752 | 0.5575 | 0.5100 | 0.5612 |
| 108th | 0.6973 | 0.8010 | 0.8300 | 0.8465 | 0.6133 | 0.8041 | 0.8930 | 0.3683 | 0.2877 | 0.4160 | 0.3049 |
| 109th | 0.6828 | 0.8078 | 0.8090 | 0.8723 | 0.5439 | 0.8577 | 0.8800 | 0.3366 | 0.2726 | 0.3710 | 0.3126 |
| 110th | 0.5848 | 0.3657 | 0.2250 | 0.2931 | 0.5337 | 0.3072 | 0.2680 | 0.7327 | 0.7443 | 0.7430 | 0.7029 |
| 111th | 0.6210 | 0.4480 | 0.4600 | 0.4267 | 0.4337 | 0.4021 | 0.3580 | 0.6644 | 0.8679 | 0.6810 | 0.7437 |
| 112th | 0.6042 | 0.4176 | 0.2980 | 0.3693 | 0.6122 | 0.2887 | 0.2910 | 0.7515 | 0.6811 | 0.7400 | 0.7505 |
| 113th | 0.6198 | 0.4922 | 0.4540 | 0.4535 | 0.5163 | 0.4856 | 0.4780 | 0.5218 | 0.5953 | 0.6330 | 0.6893 |

| Training Session | Rate Correct by Session of Congress, Naïve Bayes Classifier | | | | | | | | | |
|---|---|---|---|---|---|---|---|---|---|---|
| | All | 104th | 105th | 106th | 107th | 108th | 109th | 110th | 111th | 112th | 113th |
| All | 0.7275 | 0.7157 | 0.7400 | 0.7327 | 0.7755 | 0.8041 | 0.8600 | 0.7426 | 0.6698 | 0.6400 | 0.6019 |
| 104th | 0.7202 | 0.6765 | 0.7540 | 0.6554 | 0.6612 | 0.6938 | 0.5870 | 0.6218 | 0.5132 | 0.6170 | 0.5184 |
| 105th | 0.7708 | 0.7147 | 0.5800 | 0.7832 | 0.6816 | 0.7866 | 0.7440 | 0.6178 | 0.6009 | 0.6050 | 0.5544 |
| 106th | 0.7571 | 0.7304 | 0.7880 | 0.6436 | 0.7327 | 0.7237 | 0.7290 | 0.6109 | 0.5443 | 0.5940 | 0.4796 |
| 107th | 0.7222 | 0.5971 | 0.6100 | 0.7129 | 0.5714 | 0.6959 | 0.6600 | 0.6564 | 0.5623 | 0.6430 | 0.5223 |
| 108th | 0.7284 | 0.6284 | 0.5800 | 0.6970 | 0.6939 | 0.6907 | 0.8010 | 0.6574 | 0.5425 | 0.5940 | 0.5194 |
| 109th | 0.7348 | 0.6500 | 0.5410 | 0.6822 | 0.6255 | 0.8093 | 0.7400 | 0.7119 | 0.5500 | 0.6180 | 0.5350 |
| 110th | 0.7395 | 0.6343 | 0.5690 | 0.6069 | 0.6714 | 0.6814 | 0.7660 | 0.6634 | 0.6377 | 0.6760 | 0.5612 |
| 111th | 0.7239 | 0.5108 | 0.4900 | 0.5416 | 0.5592 | 0.5938 | 0.7160 | 0.7644 | 0.6415 | 0.7750 | 0.6816 |
| 112th | 0.7158 | 0.5627 | 0.5120 | 0.5327 | 0.6000 | 0.5660 | 0.5890 | 0.6663 | 0.7774 | 0.6300 | 0.7398 |
| 113th | 0.6758 | 0.5627 | 0.5510 | 0.5010 | 0.5418 | 0.4608 | 0.5450 | 0.5545 | 0.7868 | 0.6960 | 0.6699 |

| Training Session | Rate Correct by Session of Congress, Support Vector Machine | | | | | | | | | |
|---|---|---|---|---|---|---|---|---|---|---|
| | All | 104th | 105th | 106th | 107th | 108th | 109th | 110th | 111th | 112th | 113th |
| All | 0.8999 | 0.8627 | 0.9100 | 0.9406 | 0.9490 | 0.9381 | 0.9800 | 0.9307 | 0.8302 | 0.8600 | 0.8058 |
| 104th | 0.7351 | 0.7059 | 0.7460 | 0.7139 | 0.6439 | 0.6711 | 0.7060 | 0.6535 | 0.4755 | 0.5390 | 0.5049 |
| 105th | 0.7625 | 0.8245 | 0.7100 | 0.7891 | 0.5929 | 0.7247 | 0.7900 | 0.6802 | 0.5179 | 0.5940 | 0.4883 |
| 106th | 0.7611 | 0.7549 | 0.7730 | 0.7327 | 0.7224 | 0.7165 | 0.7310 | 0.6277 | 0.5264 | 0.5800 | 0.5184 |
| 107th | 0.8038 | 0.6686 | 0.6610 | 0.7812 | 0.7449 | 0.8175 | 0.8220 | 0.7861 | 0.6123 | 0.6720 | 0.5699 |
| 108th | 0.7574 | 0.7206 | 0.6400 | 0.7257 | 0.7408 | 0.7526 | 0.6500 | 0.7416 | 0.5283 | 0.5820 | 0.5544 |
| 109th | 0.7849 | 0.6804 | 0.6330 | 0.7188 | 0.7092 | 0.8567 | 0.8700 | 0.7693 | 0.6330 | 0.6270 | 0.5204 |
| 110th | 0.7968 | 0.6686 | 0.6060 | 0.6168 | 0.7612 | 0.7588 | 0.8830 | 0.7426 | 0.7179 | 0.6740 | 0.5825 |
| 111th | 0.6725 | 0.4941 | 0.5280 | 0.4812 | 0.5439 | 0.5093 | 0.4870 | 0.6149 | 0.7358 | 0.6710 | 0.6903 |
| 112th | 0.8141 | 0.6363 | 0.6640 | 0.7000 | 0.7408 | 0.6619 | 0.6710 | 0.8307 | 0.8283 | 0.7500 | 0.7903 |
| 113th | 0.7197 | 0.5373 | 0.6190 | 0.5594 | 0.6031 | 0.5000 | 0.4990 | 0.6426 | 0.8019 | 0.7550 | 0.7184 |

| Training Session | Rate Correct by Session of Congress, Lasso Classifier | | | | | | | | | |
|---|---|---|---|---|---|---|---|---|---|---|
| | All | 104th | 105th | 106th | 107th | 108th | 109th | 110th | 111th | 112th | 113th |
| All   | 0.9832 | 0.9804 | 0.9900 | 1.0000 | 1.0000 | 0.9897 | 1.0000 | 0.9901 | 0.9528 | 0.9800 | 0.9515 |
| 104th | 0.6947 | 0.9706 | 0.7800 | 0.8713 | 0.6633 | 0.8144 | 0.7900 | 0.4851 | 0.5283 | 0.6300 | 0.4272 |
| 105th | 0.6442 | 0.8333 | 1.0000 | 0.9109 | 0.5510 | 0.8660 | 0.7700 | 0.3069 | 0.5094 | 0.3600 | 0.3495 |
| 106th | 0.7007 | 0.8039 | 0.8900 | 0.9901 | 0.7143 | 0.7423 | 0.8800 | 0.4257 | 0.5283 | 0.5100 | 0.5437 |
| 107th | 0.7334 | 0.6667 | 0.6700 | 0.7624 | 0.9490 | 0.8144 | 0.7700 | 0.7525 | 0.6792 | 0.7200 | 0.5631 |
| 108th | 0.6670 | 0.6961 | 0.8000 | 0.8317 | 0.6429 | 0.9588 | 0.9600 | 0.4455 | 0.4717 | 0.4800 | 0.4175 |
| 109th | 0.6858 | 0.7451 | 0.7100 | 0.7426 | 0.6735 | 0.8660 | 0.9900 | 0.6535 | 0.4434 | 0.5400 | 0.5146 |
| 110th | 0.6660 | 0.5098 | 0.5100 | 0.5842 | 0.7755 | 0.4742 | 0.5800 | 0.9901 | 0.7642 | 0.7700 | 0.6990 |
| 111th | 0.6432 | 0.6275 | 0.6000 | 0.5941 | 0.5510 | 0.5258 | 0.4700 | 0.7030 | 0.9340 | 0.6900 | 0.7087 |
| 112th | 0.7255 | 0.6176 | 0.6300 | 0.7030 | 0.7653 | 0.5258 | 0.6000 | 0.8515 | 0.7642 | 0.9800 | 0.8058 |
| 113th | 0.6214 | 0.4902 | 0.5000 | 0.5347 | 0.5816 | 0.4021 | 0.3300 | 0.7327 | 0.8679 | 0.8200 | 0.9223 |

Source: Own calculations using data from Laurendale and Herzog (2016), Lewis et al. (2017).

**Table 3**: Accuracy Rate of General and Congress-Specific Classification Using DWN1 Data

| Training Session | Rate Correct by Session of Congress, Tree Classifier | | | | | | | | | |
|---|---|---|---|---|---|---|---|---|---|---|
| | All | 104th | 105th | 106th | 107th | 108th | 109th | 110th | 111th | 112th | 113th |
| All   | 0.7228 | 0.6733 | 0.6700 | 0.7400 | 0.7551 | 0.7526 | 0.7475 | 0.7100 | 0.7358 | 0.7576 | 0.6893 |
| 104th | 0.6854 | 0.8317 | 0.8500 | 0.8620 | 0.5020 | 0.7918 | 0.7980 | 0.2560 | 0.3462 | 0.4071 | 0.3806 |
| 105th | 0.6495 | 0.7960 | 0.8600 | 0.8700 | 0.4633 | 0.8072 | 0.8253 | 0.2340 | 0.3094 | 0.2545 | 0.3010 |
| 106th | 0.6776 | 0.7891 | 0.8420 | 0.8400 | 0.5673 | 0.8247 | 0.8303 | 0.2910 | 0.2472 | 0.4283 | 0.3883 |
| 107th | 0.6827 | 0.5871 | 0.5960 | 0.6110 | 0.5714 | 0.6155 | 0.6303 | 0.5680 | 0.5585 | 0.5273 | 0.5602 |
| 108th | 0.6562 | 0.7168 | 0.7760 | 0.7880 | 0.5459 | 0.6907 | 0.8404 | 0.3320 | 0.3311 | 0.3515 | 0.3078 |
| 109th | 0.6727 | 0.7861 | 0.8229 | 0.8510 | 0.5347 | 0.8278 | 0.8586 | 0.3620 | 0.3198 | 0.3414 | 0.2767 |
| 110th | 0.5579 | 0.2891 | 0.1860 | 0.2100 | 0.5224 | 0.2608 | 0.2455 | 0.7800 | 0.7264 | 0.7758 | 0.7087 |
| 111th | 0.5979 | 0.3802 | 0.3560 | 0.3510 | 0.4735 | 0.4454 | 0.4212 | 0.6290 | 0.8679 | 0.6596 | 0.701 |
| 112th | 0.5861 | 0.4158 | 0.2840 | 0.2960 | 0.6031 | 0.2691 | 0.2626 | 0.7570 | 0.6943 | 0.7677 | 0.7301 |
| 113th | 0.6235 | 0.4891 | 0.5030 | 0.5080 | 0.5235 | 0.5031 | 0.4788 | 0.4970 | 0.5642 | 0.6242 | 0.7282 |

| Training Session | Rate Correct by Session of Congress, Support Vector Machine | | | | | | | | | |
|---|---|---|---|---|---|---|---|---|---|---|
| | All | 104th | 105th | 106th | 107th | 108th | 109th | 110th | 111th | 112th | 113th |
| All   | 0.7009 | 0.7822 | 0.6600 | 0.7000 | 0.6224 | 0.7526 | 0.6869 | 0.6900 | 0.6509 | 0.7980 | 0.6699 |
| 104th | 0.6924 | 0.6139 | 0.6240 | 0.6340 | 0.6286 | 0.6753 | 0.6899 | 0.6470 | 0.5047 | 0.5364 | 0.4631 |
| 105th | 0.7940 | 0.8515 | 0.8800 | 0.8590 | 0.6449 | 0.7433 | 0.6737 | 0.5540 | 0.6991 | 0.6404 | 0.6010 |
| 106th | 0.8019 | 0.8000 | 0.8550 | 0.7600 | 0.7398 | 0.7722 | 0.7838 | 0.5640 | 0.5811 | 0.6475 | 0.5505 |
| 107th | 0.7738 | 0.5931 | 0.6400 | 0.6960 | 0.6633 | 0.7330 | 0.7404 | 0.7400 | 0.6783 | 0.6556 | 0.6019 |
| 108th | 0.7503 | 0.7614 | 0.6850 | 0.7270 | 0.6408 | 0.7320 | 0.8535 | 0.6110 | 0.5047 | 0.5697 | 0.5117 |
| 109th | 0.7783 | 0.7139 | 0.6730 | 0.6900 | 0.7306 | 0.8526 | 0.8283 | 0.7880 | 0.5208 | 0.5990 | 0.5146 |
| 110th | 0.7409 | 0.5356 | 0.5290 | 0.5110 | 0.6786 | 0.6474 | 0.7242 | 0.8100 | 0.7349 | 0.7091 | 0.6049 |
| 111th | 0.7328 | 0.5802 | 0.5950 | 0.5390 | 0.6357 | 0.5010 | 0.5465 | 0.7620 | 0.7547 | 0.7556 | 0.7223 |
| 112th | 0.7620 | 0.6396 | 0.5070 | 0.6020 | 0.6878 | 0.5856 | 0.6545 | 0.7440 | 0.8047 | 0.798 | 0.6971 |
| 113th | 0.7442 | 0.6594 | 0.6030 | 0.6650 | 0.5673 | 0.6340 | 0.5758 | 0.5940 | 0.6142 | 0.8071 | 0.7087 |

| Training Session | Rate Correct by Session of Congress, Lasso Classifier | | | | | | | | | |
|---|---|---|---|---|---|---|---|---|---|---|
| | All | 104th | 105th | 106th | 107th | 108th | 109th | 110th | 111th | 112th | 113th |
| All   | 0.9980 | 1.0000 | 1.0000 | 0.9900 | 0.9898 | 1.0000 | 1.0000 | 1.0000 | 1.0000 | 1.0000 | 1.0000 |
| 104th | 0.9940 | 1.0000 | 1.0000 | 1.0000 | 0.9898 | 0.9897 | 0.9899 | 0.9900 | 1.0000 | 0.9899 | 0.9903 |
| 105th | 1.0000 | 1.0000 | 1.0000 | 1.0000 | 1.0000 | 1.0000 | 1.0000 | 1.0000 | 1.0000 | 1.0000 | 1.0000 |
| 106th | 0.9442 | 0.9900 | 0.9900 | 1.0000 | 1.0000 | 0.7312 | 0.8788 | 0.8600 | 1.0000 | 0.9899 | 0.9903 |
| 107th | 0.9950 | 1.0000 | 1.0000 | 1.0000 | 1.0000 | 1.0000 | 1.0000 | 0.9900 | 0.9811 | 1.0000 | 0.9806 |
| 108th | 1.0000 | 1.0000 | 1.0000 | 1.0000 | 1.0000 | 1.0000 | 1.0000 | 1.0000 | 1.0000 | 1.0000 | 1.0000 |
| 109th | 1.0000 | 1.0000 | 1.0000 | 1.0000 | 1.0000 | 1.0000 | 1.0000 | 1.0000 | 1.0000 | 1.0000 | 1.0000 |
| 110th | 0.9980 | 0.9901 | 0.9900 | 1.0000 | 1.0000 | 1.0000 | 1.0000 | 1.0000 | 1.0000 | 1.0000 | 1.0000 |
| 111th | 1.0000 | 1.0000 | 1.0000 | 1.0000 | 1.0000 | 1.0000 | 1.0000 | 1.0000 | 1.0000 | 1.0000 | 1.0000 |
| 112th | 1.0000 | 1.0000 | 1.0000 | 1.0000 | 1.0000 | 1.0000 | 1.0000 | 1.0000 | 1.0000 | 1.0000 | 1.0000 |
| 113th | 1.0000 | 1.0000 | 1.0000 | 1.0000 | 1.0000 | 1.0000 | 1.0000 | 1.0000 | 1.0000 | 1.0000 | 1.0000 |

Source: Own calculations using data from Laurendale and Herzog (2016), Lewis et al. (2017).